\documentstyle[12pt,aasms4]{article}

\def\clock{\count0=\time \divide\count0 by 60
     \count1=\count0 \multiply\count1 by -60 \advance\count1 by \time
     \number\count0:\ifnum\count1<10{0\number\count1}\else\number\count1\fi}

\begin{document}

\title{Brief Note: Analytical Fit to the Luminosity Distance for Flat 
Cosmologies with a Cosmological Constant}
\author{Ue-Li Pen}
\affil{Harvard-Smithsonian Center
for Astrophysics, 60 Garden St., Cambridge, MA 02138}
\newcommand{\etal}{{\it et al.}}
\newcommand{\be}{\begin{equation}}
\newcommand{\ee}{\end{equation}}
\begin{abstract}
We present a fitting formula for the luminosity and angular diameter
distances in cosmological models with pressureless matter, a
cosmological constant and zero spatial curvature.  The formula has a
relative error of less than $0.4\%$ for $0.2<\Omega_0<1$ for any
redshift, and a global relative error of less than $4\%$ for any choice of
parameters.
\end{abstract}


\section{Introduction}

It is presently fashionable to consider spatially flat cosmological
models with a cosmological constant (Ostriker and Steinhardt 1995).
Several projects are underway to measure the magnitude of the
cosmological constant, usually based on the luminosity-distance
(Perlmutter \etal\ 1997), the angular-diameter distance (Pen 1997), or
volume-distance (Kochanek 1996).  Unfortunately, these distances are
only expressible in terms of Elliptic functions (Eisenstein 1997).  In
order to simplify the repeated computation of difficult transcendental
functions or numerical integrals, 
we present a fitting formula with the following properties:
\begin{enumerate}
\item it is exact for $\Omega_0\longrightarrow 1^-$ and 
$\Omega_0\longrightarrow 0^+$ at all redshifts.
\item The relative error tends to zero as $z \rightarrow \infty$
for any value of $\Omega_0$.
\item For the range $0.2\le\Omega_0\le 1$, the relative error is less
than $0.4\%$.
\item For any choice of parameters, the relative error is always less
than $4\%$.
\end{enumerate}
Without further ado, the luminosity distance is given as:
\begin{eqnarray}
d_L &=&
\frac{c}{H_0}(1+z)\left[\eta(1,\Omega_0)-\eta(\frac{1}{1+z},\Omega_0)\right]
\nonumber\\
\eta(a,\Omega_0) &=& 2\sqrt{s^3+1}
\left[\frac{1}{a^4}-0.1540\frac{s}{a^3}+0.4304\frac{s^2}{a^2}
+0.19097\frac{s^3}{a}+0.066941s^4\right]^{-\frac{1}{8}}
\nonumber\\
s^3&=&\frac{1-\Omega_0}{\Omega_0}.
\label{eqn:dl}
\end{eqnarray}
We have used the Hubble constant $H_0$, and the pressureless matter
content $\Omega_0$.  We recall
that an object of luminosity $L$ has flux $F=L/(4\pi d_L^2)$.

\section{Approximation}

\begin{figure}
\plotone{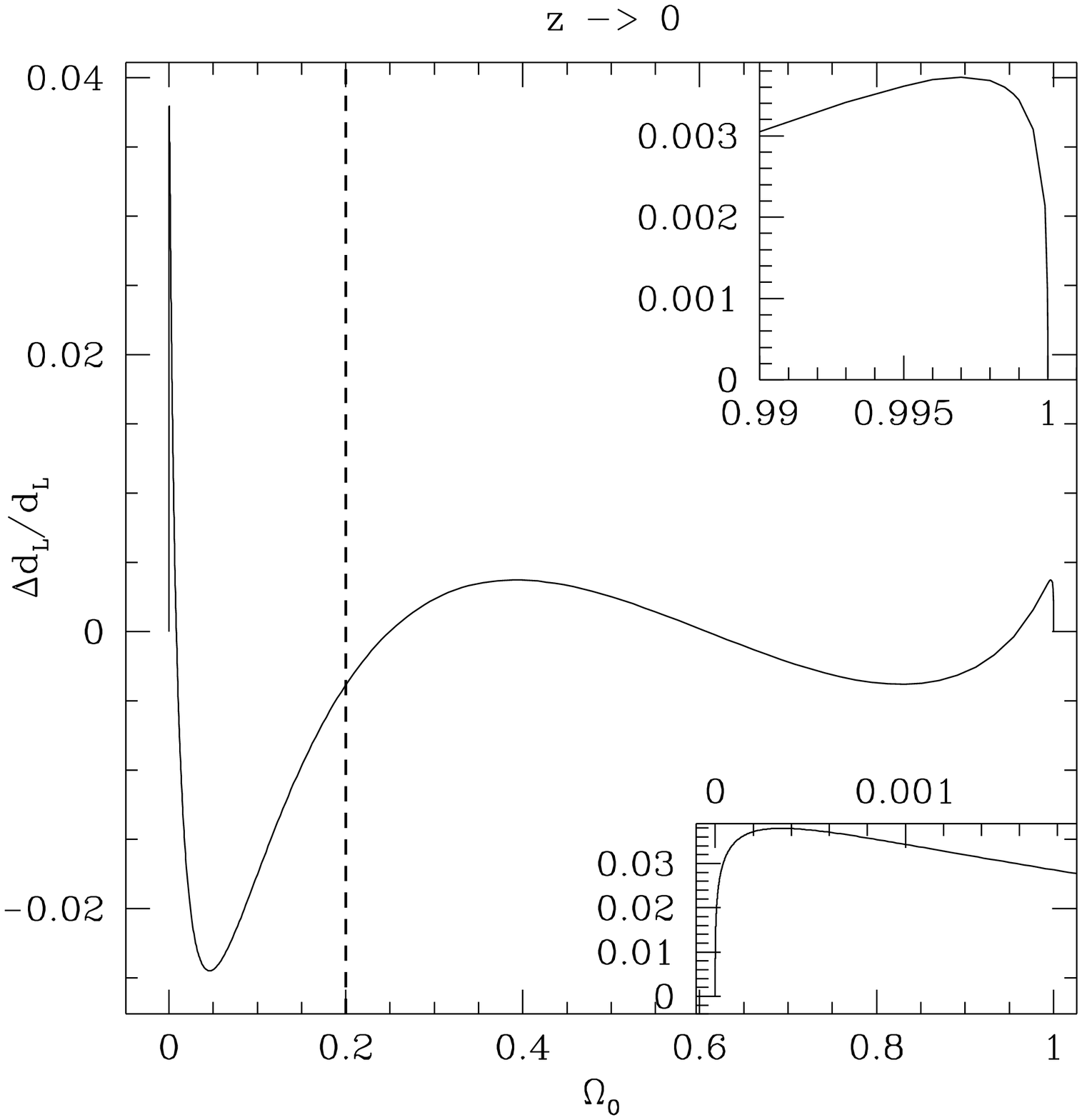}
\caption{Relative error at $z\rightarrow 0$, which dominates the global error
budget.  The minmax error was
optimized in the range $0.2\le \Omega_0 \le 1$, which is the area to the 
right of the vertical dashed curve.  The error goes to zero as a powerlaw at 
both ends of the graph, which is shown in the two insets.}
\label{fig:z0}
\end{figure}

\begin{figure}
\plotone{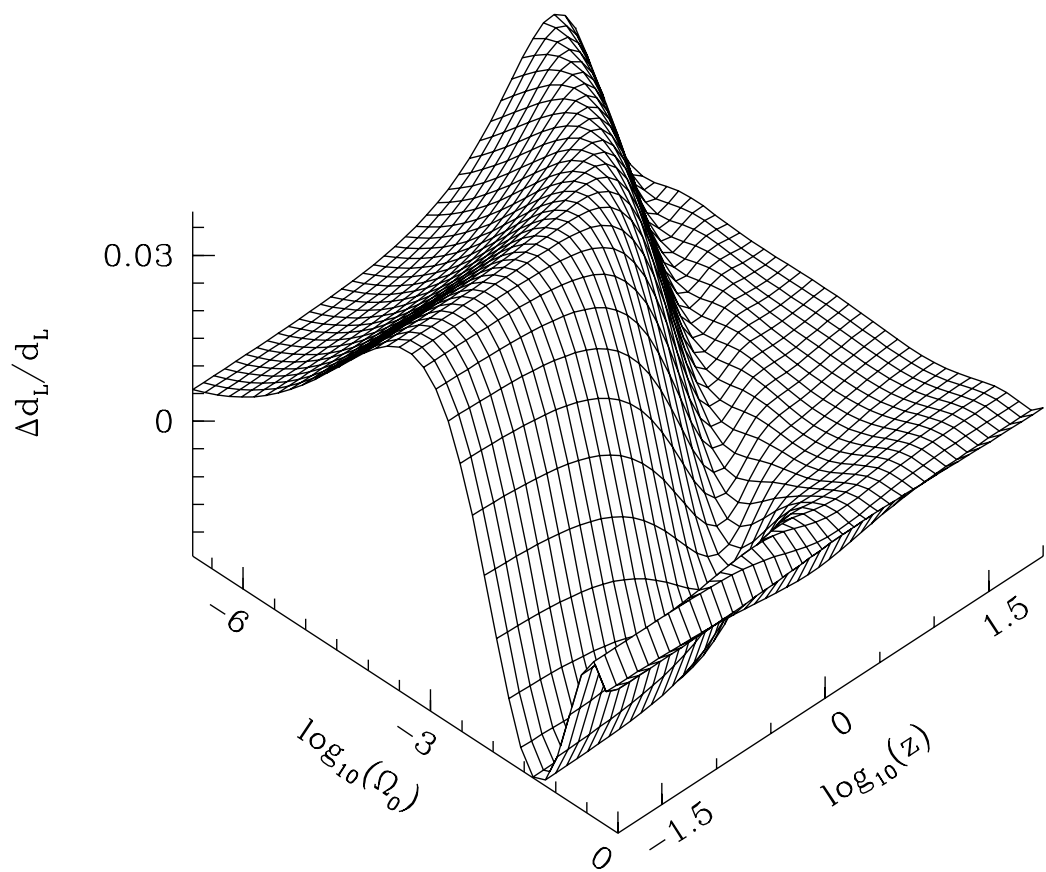}
\caption{Global error surface.  Our minmax fit minimized all errors
for $\log_{10}\Omega_0\ge -.7$, where the error is less than $0.4\%$
for all redshifts.}
\label{fig:err2d}
\end{figure}

\begin{figure}
\plotone{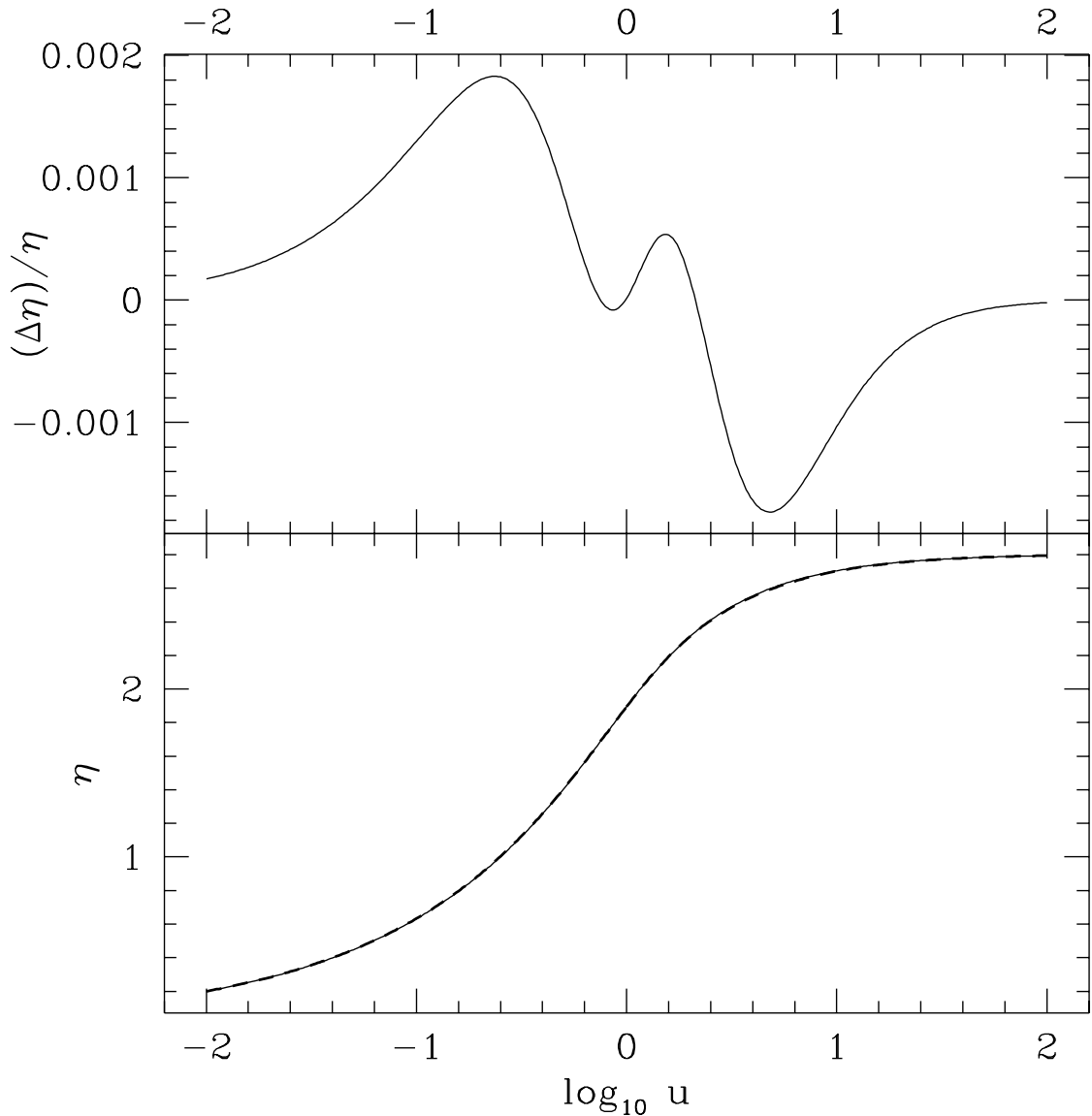}
\caption{Exact and approximate conformal time.
The lower figure shows the conformal time as a function of
$u=[(1-\Omega_0)/\Omega_0]^{1/3}/(1+z)$ for the exact integral (solid
line) and the approximation of equation (\protect\ref{eqn:dl})
(dashed).  These two lines are practically indiscernible.
The upper figure shows the relative error $(\tilde\eta-\eta)/\eta$.}
\label{fig:eta}
\end{figure}

Spatial flatness allows us to rewrite the Friedman-Robertson-Walker
metric as a conformally flat spacetime
\be
ds^2=a^2(-d\eta^2+dr^2+r^2d\Omega).
\ee
$\eta$ is the conformal time, and $r$ is the comoving distance.  We
have set the speed of light $c=1$.  The
scale factor $a$ can be normalized in terms of the redshift $a\equiv
1/(1+z)$.  The Friedman equations determine 
\be
\left(\frac{da}{d\eta}\right)^2=a\Omega_0+a^4\Omega_\Lambda
\label{eqn:frw}
\ee
where spatial flatness requires $\Omega_0+\Omega_\Lambda=1$.  A change
of variables to $u=as$ where $s^3=(1-\Omega_0)/\Omega_0$ allows us to
express (\ref{eqn:frw}) parameter free
\be
\eta = \sqrt{\frac{s^3+1}{s}}\int_0^{as} \frac{du}{\sqrt{u^4+u}}.
\label{eqn:etau}
\ee
We can asymptotically approximate (\ref{eqn:etau}) as
\be
\eta_1=2\sqrt{\frac{s^3+1}{s}}
\left[u^{-n}+n(2X)^{2n+1}u^{-1}+(2X)^{2n}\right]^\frac{-1}{2n}
\label{eqn:eta1}
\ee
where $X\equiv (\int_0^\infty du/\sqrt{u^4+u})^{-1}=3\sqrt{\pi}/
(\Gamma[\frac{1}{6}]\Gamma[\frac{1}{3}])\simeq 0.3566$. $n$
is a free parameter, and at this stage a choice of $n=3$ approximates
all distances to better than $14\%$ relative accuracy.  Equation
(\ref{eqn:eta1}) satisfies the following conditions: 1. it converges
to (\ref{eqn:etau}) as $u\rightarrow 0$ and
$u\rightarrow\infty$. 2. Its derivative converges to the derivative of
$\eta$ as $u\rightarrow \infty$.  To improve on (\ref{eqn:eta1}) we
consider a polynomial expansion of $u^{-1}$ in the denominator with
two free parameters by setting $n=4$:
\be
\tilde\eta=2\sqrt{\frac{s^3+1}{s}}
\left[u^{-4}+c_1u^{-3}+c_2u^{-2}+4(2X)^{9}u^{-1}
+(2X)^{8}\right]^{-1/8}.
\label{eta2}
\ee
We will now choose the coefficients $c_1,c_2$ to minimize the relative
error in the approximate luminosity distance $\tilde d_L$,
\be
e_\infty \equiv \|(\tilde d_L-d_L)/d_L\|_\infty
\label{eqn:err}
\ee  
where the subscript $\infty$ indicates the infinity norm, i.e. the
maximal value over the domain.
The error in (\ref{eqn:err}) tends to be dominated by $z=0$ (see for
example Figure \ref{fig:err2d}), for which we can express
\be
e_\infty = \left\| \sqrt{u^4+u}\frac{d\tilde\eta}{du} - 1 \right\|_\infty.
\label{eqn:erru}
\ee
Globally optimizing (\ref{eqn:erru}) allows us to reduce the error to
about $2\%$.  But we choose the following trade-off for current
cosmological parameters:  We want to minimize (\ref{eqn:erru}) over
the range $0.2\le \Omega_0\le 1$, which covers the popularly
considered parameter space.  In that range, we find through a
non-linear equation solver that $c_1=-0.1540$ and $c_2=0.4304$.  This
allows us to trade the global error of $2\%$ for a global error of
$4\%$, while reducing the error in the range of interest to $0.4\%$.
The global error surface plot is shown in Figure \ref{fig:err2d}.
The error at $z\rightarrow 0$ is shown in Figure \ref{fig:z0}.
At small $z$, any errors in $\tilde\eta$ are amplified, even though the global
errors in $\eta(z)$ are generally significantly smaller.
Figure (\ref{fig:eta}) shows the fit for the conformal time $\tilde\eta$
and its residual, which is accurate to $0.2\%$ globally.  

One can always express the luminosity distance as a power series
expansion around $z=0$ using Equations (\ref{eqn:dl}) and
(\ref{eqn:etau}).  For a flat universe, one obtains
\be
\frac{H_0}{c}d_L = z + \left(1-\frac{3}{4}\Omega_0\right) z^2
+(9\Omega_0-10)\frac{\Omega_0}{8}z^3 + \cdots
\label{eqn:series}
\ee
The series converges only slowly for $z\sim 1$: even when expanded to
sixth order in $z$, the maximal relative
error in the interval $0<z<1$ and $0.2<\Omega_0<1$ using the series
expansion in (\ref{eqn:series}) is 37\%.

\section{Conclusion}

We have presented a simple algebraic approximation to the luminosity
distance $d_L$ and the proper angular diameter distance
$d_A=d_L/(1+z)^2$ in a flat universe with pressureless matter and a
cosmological constant.

\newpage

\end{document}